\input amstex
\magnification=1200
\documentstyle{amsppt}
\NoRunningHeads
\NoBlackBoxes
\topmatter
\title Quantum string field psychophysics of nastroenie
\endtitle
\author Denis V. Juriev\endauthor
\affil ul.Miklukho-Maklaya 20-180 Moscow 117437 Russia\linebreak
(e-mail: denis\@juriev.msk.ru)\endaffil
\date
\enddate
\endtopmatter
\document
The goal of this article is to expose some of results, which were obtained
by the author last year and which may be interesting to a general reader.
So the technical details (such as experimental data processing and their 
analysis) are omitted or reduced whereas the conceptual aspect is inforced
and clarified.

\head I. What is known [1]\endhead

\subhead 1.1. Experimental detection of interactive phenomena\endsubhead
Let us consider a natural, behavioral, social or economical system $\Cal S$.
It will be described by a set $\{\varphi\}$ of quntities, which characterize
it at any moment of time $t$ (so that $\varphi=\varphi_t$). One may suppose
that the evolution of the system is described by a differential equation 
$$\dot\varphi=\Phi(\varphi)$$
and look for the explicit form of the function $\Phi$ from the experimental
data on the system $\Cal S$. However, the function $\Phi$ may depend on time,
it means that there are some hidden parameters, which control the system
$\Cal S$ and its evolution is of the form
$$\dot\varphi=\Phi(\varphi,u),$$
where $u$ are such parameters of unknown nature. One may suspect that such 
parameters are chosen in a way to minimize some goal function $K$, which may 
be an integrodifferential functional of $\varphi_t$:
$$K=K(\left[\varphi_{\tau}\right]_{\tau\le t})$$
(such integrodifferential dependence will be briefly notated as 
$K=K([\varphi])$ below). More generally, the parameters $u$ may be divided
on parts $u=(u_1,\ldots,u_n)$ and each part $u_i$ has its own goal function
$K_i$. However, this hypothesis may be confirmed by the experiment very 
rarely. In the most cases the choice of parameters $u$ will seem accidental
or even random. Nevertheless, one may suspect that the controls $u_i$ are 
{\sl interactive}, it means that they are the couplings of the pure controls 
$u_i^\circ$ with the {\sl unknown or incompletely known\/} feedbacks:
$$u_i=u_i(u_i^\circ,[\varphi])$$
and each pure control has its own goal function $K_i$. Thus, it is
suspected that the system $\Cal S$ realizes an {\sl interactive game}.
There are several ways to define the pure controls $u_i^\circ$. One of them
is the integrodifferential filtration of the controls $u_i$:
$$u^\circ_i=F_i([u_i],[\varphi]).$$
To verify the formulated hypothesis and to find the explicit form of the
convenient filtrations $F_i$ and goal functions $K_i$ one should use the
theory of interactive games, which supplies us by the predictions of the
game, and compare the predictions with the real history of the game for
any considered $F_i$ and $K_i$ and choose such filtrations and goal functions,
which describe the reality better. One may suspect that the dependence of
$u_i$ on $\varphi$ is purely differential for simplicity or to introduce the
so-called {\sl intention fields}, which allow to consider any interactive
game as differential. Moreover, one may suppose that
$$u_i=u_i(u_i^\circ,\varphi)$$
and apply the elaborated procedures of {\sl a posteriori\/} analysis and
predictions to the system.

In many cases this simple algorithm effectively unravels the hidden 
interactivity of a complex system. However, more sophisticated psychophysical
procedures exist.

Below we shall consider the complex systems $\Cal S$, which have been yet
represented as the $n$-person interactive games by the procedure described
above. 

\subhead 1.2. Functional analysis of interactive phenomena\endsubhead
To perform an analysis of the interactive control let us note that often for 
the $n$-person interactive game the interactive controls 
$u_i=u_i(u_i^\circ,[\varphi])$ may be represented in the form 
$$u_i=u_i(u_i^\circ,[\varphi];\varepsilon_i),$$
where the dependence of the interactive controls on the arguments
$u_i^\circ$, $[\varphi]$ and $\varepsilon_i$ is known but the 
$\varepsilon$-parameters $\varepsilon_i$ are the unknown or incompletely
known functions of $u_i^\circ$, $[\varepsilon]$. Such representation is
very useful in the theory of interactive games and is called the 
{\sl $\varepsilon$-representation}. 

One may regard $\varepsilon$-parameters as new magnitudes, which characterize
the system, and apply the algorithm of the unraveling of interactivity to
them. Note that $\varepsilon$-parameters are of an existential nature 
depending as on the states $\varphi$ of the system $\Cal S$ as on the
controls. 

The $\varepsilon$-parameters are useful for the functional analysis of
the interactive controls described below.

First of all, let us consider new integrodifferential filtrations $V_\alpha$:
$$v^\circ_\alpha=V_\alpha([\varepsilon],[\varphi]),$$
where $\varepsilon=(\varepsilon_1,\ldots,\varepsilon_n)$. 
Second, we shall suppose that the $\varepsilon$-parameters are expressed via 
the new controls $v^\circ_\alpha$, which will be called {\it desires:}
$$\varepsilon_i=\varepsilon_i(v^\circ_1,\ldots,v^\circ_m,[\varphi])$$
and the least have the goal functions $L_\alpha$. The procedure of unraveling
of interactivity specifies as the filtrations $V_\alpha$ as the goal functions 
$L_\alpha$.

\subhead 1.3. SD-transform and SD-pairs\endsubhead
The interesting feature of the proposed description (which will be called the
{\it S-picture}\/) of an interactive system $\Cal S$ is that it contains as 
the real (usually personal) subjects with the pure controls $u_i$ as the 
impersonal desires $v_\alpha$. The least are interpreted as certain 
perturbations of the first so the subjects act in the system by the 
interactive controls $u_i$ whereas the desires are hidden in their actions. 

One is able to construct the dual picture (the {\sl D-picture\/}),
where the desires act in the system $\Cal S$ interactively and the
pure controls of the real subjects are hidden in their actions.
Precisely, the evolution of the system is governed by the equations
$$\dot\varphi=\tilde\Phi(\varphi,v),$$
where $v=(v_1,\ldots,v_m)$ are the $\varepsilon$-represented interactive 
desires:
$$v_\alpha=v_\alpha(v^\circ_\alpha,[\varphi];\tilde\varepsilon_\alpha)$$
and the $\varepsilon$-parameters $\tilde\varepsilon$ are the unknown or
incompletely known functions of the states $[\varphi]$ and the pure
controls $u_i^\circ$.

D-picture is convenient for a description of systems $\Cal S$ with a
variable number of acting persons. Addition of a new person does not
make any influence on the evolution equations, a subsidiary term to
the $\varepsilon$-parameters should be added only.

The transition from the S-picture to the D-picture is called the
{\it SD-transform}. The {\it SD-pair\/} is defined by the evolution
equations in the system $\Cal S$ of the form
$$\dot\varphi=\Phi(\varphi,u)=\tilde\Phi(\varphi,v),$$
where $u=(u_1,\ldots,u_n)$, $v=(v_1,\ldots,v_m)$, 
$$\aligned
u_i=&u_i(u_i^\circ,[\varphi];\varepsilon_i)\\
v_\alpha=&v_\alpha(v^\circ_\alpha,[\varphi];\tilde\varepsilon_\alpha)
\endaligned$$
and the $\varepsilon$-parameters $\varepsilon=(\varepsilon_1,\ldots,
\varepsilon_n)$ and $\tilde\varepsilon=(\tilde\varepsilon_1,\ldots,
\tilde\varepsilon_m)$ are the unknown or incompletely known functions of
$[\varphi]$ and $v^\circ=(v^\circ_1,\ldots,v^\circ_m)$ or
$u^\circ=(u^\circ_1,\ldots,u^\circ_n)$, respectively. 

Note that the S-picture and the D-picture may be regarded as complementary
in the N.Bohr sense. Both descriptions of the system $\Cal S$ can not be 
applied to it simultaneously during its analysis, however, they are compatible 
and the structure of SD-pair is a manifestation of their compatibility.
The choice of a picture is an action of our {\it attention:\/} it is
concentrated on the personal subjects in S-picture {\it (the self-conscious 
attention)\/} whereas it is concentrated on the impersonal desires in 
D-picture {\it (the creative attention)}.

\subhead 1.4. Verbalization of SD-pairs and the transpersonal synlinguism.
Nastroenie\endsubhead The main problem is to interrelate the S- and
D-pictures of the system $\Cal S$. One way is a {\it verbalization\/} of
SD-pairs. Let us remind a definition of the verbalizable interactive game.
 
An interactive game of the form
$$\dot\varphi=\Phi(\varphi,u)$$
with $\varepsilon$--represented couplings of feedbacks 
$$u_i=u_i(u^\circ_i,[\varphi];\varepsilon_i)$$
is called {\it verbalizable\/} if there exist {\sl a posteriori\/}
partition $t_0\!<\!t_1\!<\!t_2\!<\!\ldots\!<\!t_n\!<\!\ldots$ and the 
integrodifferential functionals
$$\aligned
\omega_n&(\vec\varepsilon(\tau),\varphi(\tau)|
t_{n-1}\!\leqslant\!\tau\!\leqslant\!t_n),\\
u^*_n&(u^\circ(\tau),\varphi(\tau)|
t_{n-1}\!\leqslant\!\tau\!\leqslant\!t_n)
\endaligned$$ 
such that
$$\omega_n=\Omega(\omega_{n-1},u^*_n;\varphi(\tau)|
t_{n-1}\!\leqslant\!\tau\!\leqslant\!t_n),$$
quantities $\omega_n$ are called the {\it words}. 

Let us now consider the SD-pair and suppose that both S- and D-pictures are
verbalizable with the {\sl same\/} $\omega_n$. The fact that $\omega_n$ are
the same for both S- and D-pictures is called the {\it transpersonal
synlinguism}. One may characterize it poetically by the phrase that {\sl
``the speech of real subjects is resulted in the same text as a whisper of
the impersonal desires''}. The existential character of the transpersonal
synlinguism should be stressed. Really it is not derived from the fact that 
the objective states $\varphi$ of the system $\Cal S$ are the same in the 
S- and D-pictures. The transpersonal synlinguism interrelates the different 
$\varepsilon$-parameters of existential nature in both pictures.

The transpersonal synlinguism is involved into psychophysical nature of mutual
understanding of the independent subjects of a dialogue communication. In this
situation it allows to identify the personal interpretations with the
impersonal ones, unraveling the role of impersonal desires as bearers of the
objective sense and its dynamics.

The words $\omega_n$ in the transpersonal synlinguism are interrelated by
the recurrent formulas
$$\omega_n=\Omega(\omega_{n-1},w^*_n;\varphi(\tau)|
t_{n-1}\!\leqslant\!\tau\!\leqslant\!t_n),$$
the parameter $w_n^*$ will be called {\it nastroenie}\footnote"*"
{``Nastroenie'' is the Russian for English ``mood'' or ``humour'' and
for French ``humeur'' or ``disposition'', however, its meaning has some 
nuances. First, its root is associated with Russian verb, which may be
translated into English as ``to build'', ``to construct'' and ``to create'',
thus, expressing the active character of ``nastroenie'' in contrast to
completely passive ``disposition''. The root ``stro\v\i '' being treated as 
a noun means ``order'', ``system'' and also ``harmony''. The related adjective
``stro\v\i ny\v\i '' means ``well-proportioned'', ``well-composed'', ``shapely''.
Second, the nearest Russian word of the same root (``nastro\v\i ka'') should
be translated as ``tuning'' into English, thus, introducing additional
harmonic, melodious overtones into the meaning. These facts expain the choice
of the Russian term.\newline}. The sequence $\{w_n\}$
may be correlated as with $u(t)$ as with $v(t)$, the first type of correlations
will be called the {\it impression correlations\/} and the second type will
be called the {\it inspiration correlations}.

There is a lot of various, sometimes technically very sophisticated procedures
to verbalize interactive game or SD-pair.

\subhead 1.5. The second quantization of desires\endsubhead
Intuitively it is reasonable to consider systems with a variable number
of desires. It can be done via the second quantization. 

To perform the second quantization of desires let us mention that they
are defined as the integrodifferential functionals of $\varphi$ and
$\varepsilon$ via the integrodifferential filtrations. So one is able
to define the linear space $H$ of all filtrations (regarded as classical 
fields) and a submanifold $M$ of the dual $H^*$ so that $H$ is naturally
identified with a subspace of the linear space $\Cal O(M)$ of smooth functions
on $M$. The quantized fields of desires are certain operators in the
space $\Cal O(M)$ (one is able to regard them as unbounded operators in its
certain Hilbert completion). The creation/annihilation operators are
constructed from the operators of multiplication on an element of $H\subset
\Cal O(M)$ and their conjugates.

To define the quantum dynamics one should separate the quick and slow time.
Quick time is used to make a filtration and the dynamics is realized in
slow time. Such dynamics may have a Hamiltonian form being governed by
a quantum Hamiltonian, which is usually differential operator in $\Cal O(M)$.

If $M$ coincides with the whole $H^*$ then the quadratic part of a Hamiltonian
describes a propagator of the quantum desire whereas the highest terms
correspond to the vertex structure of self-interaction of the quantum field. 
If the submanifold $M$ is nonlinear, the extraction of propagators and 
interaction vertices is not straightforward.

\subhead 1.6. Quantum string field theoretic structure of the second 
quantization of desires\endsubhead
First of all, let us mark that the functions $\varphi(\tau)$ and
$\varepsilon(\tau)$ may be regarded formally as an open string. The target 
space is a product of the spaces of states and $\varepsilon$-parameters. 

Second, let us consider a classical counterpart of the evolution of the 
integrodifferential filtration. It is natural to suspect that such evolution
is local in time, i.e. filtrations do not enlarge their support (as a time
interval) during their evolution. For instance, if the integrodifferential
filtration depends on the values of $\varphi(\tau)$, $\varepsilon(\tau)$
for $\tau\in[t_0-t_1,t_0-t_2]$ at the fixed moment $t_0$, it will depend
on the same values for $\tau\in[t-t_1,t-t_2]$ at other moments $t>t_0$.
This supposition provides the reparametrization invariance of the classical
evolution. Hence, it is reasonable to think that the quantum evolution is
also reparametrization invariant. 

Reparametrization invariance allows to apply the quantum string field 
theoretic models to the second quantization of desires. For instance, one
may use the string field actions constructed from the closed string vertices
(note that the phase space for an open string coincides with the configuration
space of a closed string) or string field theoretic nonperturbative actions.
In the least case the theoretic presence of additional ``vacua'' (minimums
of the string field action) as well as their structure is very interesting
(see below).

\head II. Questions and answers\endhead

The described picture contains some moments, which should be clarified:
\roster
\item"-" What is the origin of the complementarity of S- and D-pictures ?
Can it be described precisely ?
\item"-" What is the nature of nastroenie ? Is it possible to describe its
dynamics and teleology ?
\item"-" What is the attention ? What is its role ?
\endroster

We shall try to answer these questions below.

\subhead 3.1. Subjects as massive particles of the universal psychophysical
string field at alternative vacuum\endsubhead
Quantized desires may be treated as interacting particles at the main vacuum
of the universal psychophysical string field. However, this field may possess
an additional vacuum of very massive (and, hence, almost classical) and almost
free particles. We shall identify subjects with such particles. Because
the spectrum of particles at different vacua is a dynamical property of
the quantum field, it depends on the manner of observation. The different
choice of vacua corresponds to the complementary observations. This fact
explains why S- and D-pictures are complementary and allows to describe
the complementarity precisely if the structure of universal psychophysical
string field is known.

\subhead 3.2. The universal psychophysical string field as perturbed
free field of nastroenie\endsubhead
One may suspect that the universal psychophysical string field may be
treated as a perturbed free field so that its action is represented as
a free field action perturbed by a potential of self-interaction, which
exponentially tends to zero at infinity. The initial vacuum of free field
disappears and perturbation creates two new vacua (of subjects and of
desires). It is reasonable to think this initial free field as a field of
nastroenie. It means that instead of the discrete quantities $w_n^*$ we
shall consider the quantized field $w(t)$, whose dynamics is described
by the perturbed free field action. However, the problem of its teleology
is more difficult, claims a well-elaborated technique to be defined, which,
therefore, will not be exposed in this article.

Note that there are no arguments to suppose that the initial free field 
has a string nature, in particular, is a free string field.

\subhead 3.3. Three types of attention as three forms of internal observations
in the quantum string field psychophysics\endsubhead
Three forms to represent the universal psychophysical string field, namely,
as series in the main or the alternative vacua or as a perturbed free field,
corresponds to three forms of observation. They were defined as attentions.
Two of them are self-conscious and creative attention. The form of attention
specifies the dynamical properties of the universal psychophysical string
field that is expressed as a choice of picture. One should mention that
the attention is attributed to any subject and should be treated as a form
of intention. Therefore, we have deal with the internal observations.

\Refs
\roster
\item"[1]" Juriev D., Experimental detection of interactive phenomena and
their analysis:\linebreak math.GM/0003001; New mathematical methods for 
psychophysical filtering of experimental data and their processing: 
math.GM/0005275; Quantum string field theory and psychophysics: 
physics/0008058.
\endroster
\endRefs
\enddocument